\documentclass[preprint,preprintnumbers,nofootinbib]{revtex4}
\usepackage[dvips]{color}
\usepackage{graphicx}

\newcommand{\hif}{\mathchar`-}

\newcommand{\diag}{{\rm diag}}
\newcommand{\ev}{{\rm eV}}

\newcommand{\gev}{{\rm GeV}}
\newcommand{\tev}{{\rm TeV}}

\begin{document}

\title{$\mu\hif\tau$ symmetry in Zee-Babu model}
\author{Takeshi Araki\footnote{araki@phys.nthu.edu.tw}  and 
C.~Q.~Geng\footnote{geng@phys.nthu.edu.tw} }
\affiliation{Department of Physics, National Tsing Hua University,
Hsinchu, Taiwan 300}

\begin{abstract}
We study the Zee-Babu two-loop neutrino mass generation model and
look for a possible flavor symmetry behind the tri-bimaximal 
neutrino mixing.
We find that there probably exists the $\mu\hif\tau$ symmetry 
in the case of the normal neutrino mass hierarchy, whereas 
there may not be in the inverted hierarchy case.
We also propose a specific model based on a Froggatt-Nielsen-like 
$Z_5$ symmetry to naturally accomplish the $\mu\hif\tau$ symmetry 
on the neutrino mass matrix for the normal hierarchy case.
\end{abstract}

\maketitle

\section{Introduction}
Neutrino oscillation experiments have almost 
completely established that neutrinos have tiny masses and mix with 
each other through the Pontecorvo-Maki-Nakagawa-Sakata (PMNS) 
leptonic mixing matrix \cite{PMNS}.
From the latest global analysis of three-neutrino mixing \cite{gfit}, 
one currently has the following best fit values with $1\sigma$ errors:
\begin{eqnarray}
&&\Delta m_{21}^2 = (7.59 \pm 0.20) \times 10^{-5}\ \ev^2 , \nonumber \\
&&\Delta m_{31}^2 = 
\left\{\begin{array}{l}
(-2.36 \pm 0.11) \times 10^{-3}\ \ev^2\ \ \ {\rm for\ inverted\ hierarchy} \\ 
(+2.46 \pm 0.12) \times 10^{-3}\ \ev^2\ \ \ {\rm for\ normal\ hierarchy} 
\end{array}\right. \ ,\label{eq:best} \\
&&\sin^2\theta_{12} = 0.319 \pm 0.016,\ \ 
\sin^2\theta_{23} = 0.462^{+0.082}_{-0.050},\ \ 
\sin^2\theta_{13} = 0.0095^{+0.013}_{-0.007}\ .\nonumber
\end{eqnarray}
The data indicate the existence of, at least, two massive neutrinos with a 
very suggestive neutrino mixing matrix, that is, 
the tri-bimaximal (TB) mixing matrix \cite{TB}:
\begin{eqnarray}
V_{TB}=\frac{1}{\sqrt{6}}
 \left(\begin{array}{ccc}
   2 & \sqrt{2} & 0 \\
  -1 & \sqrt{2} & \sqrt{3} \\
   1 &-\sqrt{2} & \sqrt{3}
 \end{array}\right)\label{eq:TB}.
\end{eqnarray}
However, the standard model (SM) neither includes neutrino mass terms nor 
provides us with any explanation for the TB mixing.
Clearly, we need new physics beyond the SM.
In fact, many extensions of the SM have been proposed so far.
For instance, in the type-I \cite{type1}, type-II \cite{type2} and 
type-III \cite{type3} seesaw mechanisms, the SM is extended by introducing 
extra heavy fermions or scalars to generate neutrino masses 
suppressed by the mass scale of the heavy particles, while 
in Ref. \cite{dim5} tiny neutrino masses come from the dimension-five 
Weinberg operators.
These scenarios have been extensively studied with some flavor 
symmetries to explain the TB mixing \cite{mu-tau,A4,S4,tani-rev,FL}.

Yet another possibility of leading to tiny neutrino masses is to use 
radiative corrections.
It was first pointed out by Zee in Ref. \cite{zee} in which new scalars are 
added in the Higgs sector with neutrino masses induced at the one-loop level.
After that, a two-loop scenario called Zee-Babu model \cite{zee-babu} was 
proposed\footnote{
Other types of multi-loop scenarios have also been studied in 
Ref. \cite{variety}.
}.
In these kinds of scenarios, discussions about the neutrino phenomena 
can be much different form those of the tree level scenarios 
because the induced neutrino mass matrix elements are the products of 
the anti-symmetric Yukawa coupling and charged lepton mass.
Hence, it is non-trivial whether a flavor symmetry can play an important 
role in the radiative scenarios.
Since there is a claim that the original Zee model may not be able to 
reproduce current neutrino oscillation data \cite{antiZee}, we focus 
on the Zee-Babu two-loop model in this Letter.
We re-analyze the model and try to explain the TB pattern 
of neutrino mixings in terms of the $\mu\hif\tau$ symmetry which is the prime 
candidate of a flavor symmetry in the tree level scenarios.
Note that other phenomenological studies have been discussed in 
Refs. \cite{babu-mace,pheno}.

This Letter is organized as follows.
In Section II, we summarize the Zee-Babu model and show some  
definitions of parameters.
In Section III, we investigate the model along with the $\mu\hif\tau$ symmetry.
We propose a specific flavor model in Section IV.
Finally, we conclude our discussions in Section V.

\section{Zee-Babu Model}
\begin{figure}[t]
\begin{center}
\includegraphics*[width=0.8\textwidth]{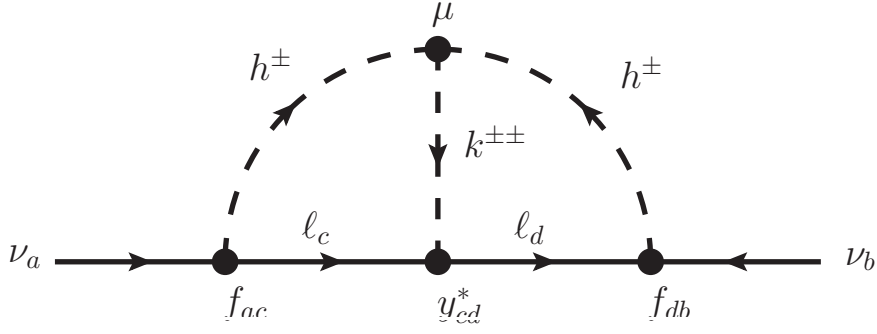}
\caption{\footnotesize
Two-loop diagram for Majorana neutrino masses. 
}\label{fig:2loop-dia}
\end{center}
\end{figure}

In addition to the SM particles, the Zee-Babu model contains two 
$SU(2)_L$ singlet new scalars: a singly charged scalar $h^\pm$ and 
doubly charged scalar $k^{\pm\pm}$.
Accordingly, new interactions appear and terms relevant to our study are
\begin{eqnarray}
{\cal L}_{ZB}=F_{ab} (L^T_a C L_b h^+ ) 
+ Y_{ab} ( \ell_{Ra}^T C \ell_{Rb} k^{++}) - \mu h^+ h^+ k^{--} + h.c.\ ,
\label{eq:zbL}
\end{eqnarray}
where $C$ is the charge conjugation matrix, $L_{a=e,\mu,\tau}$ stand for 
the left-handed $SU(2)_L$ doublet leptons and $\ell_{Ra}$ are the right-handed 
singlet charged leptons in the diagonal basis of the charged lepton mass matrix.
$F_{ab}$ and $Y_{ab}$ are $3\times 3$ complex Yukawa matrices, parametrized as
\begin{eqnarray}
F_{ab}=
 \left(\begin{array}{ccc}
 0 & f_{e\mu} & f_{e\tau} \\
 -f_{e\mu} & 0 & f_{\mu\tau} \\
 -f_{e\tau} & -f_{\mu\tau} & 0
 \end{array}\right),
\hspace{1cm}
Y_{ab}=
 \left(\begin{array}{ccc}
 y_{ee} & y_{e\mu} & y_{e\tau} \\
 y_{e\mu} & y_{\mu\mu} & y_{\mu\tau} \\
 y_{e\tau} & y_{\mu\tau} & y_{\tau\tau}
 \end{array}\right)\ .
\label{eq:FY}
\end{eqnarray}
The Majorana neutrino mass term is induced at the two-loop level as depicted 
in Fig. \ref{fig:2loop-dia} with the mass matrix given by 
\begin{eqnarray}
{\cal M}_{ab} = 8\mu (f_{ac} m_c y_{cd}^* m_d f_{db}) I \label{eq:Mzb},
\end{eqnarray}
where $m_a$ indicate the charged lepton masses and 
\begin{eqnarray}
I\simeq\frac{1}{(16\pi^2)^2}\frac{1}{M_h^2} \int^1_0 dx \int^{1-x}_0 dy
  \frac{-(1-y)}{x+\left[(M_k/M_h)^2 -1\right]y+y^2}
  \log\frac{y(1-y)}{x+(M_k/M_h)^2 y}\label{eq:I}
\end{eqnarray}
is the two-loop integral function with the masses of the new scalars, 
$M_k$ and $M_h$.
Note that Eq. (\ref{eq:I}) is simplified by neglecting the charged 
lepton masses \cite{babu-mace}. 
The elements of the neutrino mass matrix in Eq. (\ref{eq:Mzb}) 
are written as
\begin{eqnarray}
\begin{array}{l}
{\cal M}_{11} = 8\mu f_{\mu\tau}^2 
(-\tilde{f}_{e\tau}^2\omega_{\tau\tau} 
 - 2\tilde{f}_{e\mu}\tilde{f}_{e\tau}\omega_{\mu\tau} 
 - \tilde{f}_{e\mu}^2 \omega_{\mu\mu}) I, \\
{\cal M}_{22} = 8\mu f_{\mu\tau}^2 
(- \omega_{\tau\tau} + 2\tilde{f}_{e\mu}\omega_{e\tau} 
 - \tilde{f}_{e\mu}^2\omega_{ee}) I, \\
{\cal M}_{33} = 8\mu f_{\mu\tau}^2 
(- \omega_{\mu\mu} - 2\tilde{f}_{e\tau}\omega_{e\mu} 
 - \tilde{f}_{e\tau}^2\omega_{ee}) I, \\
{\cal M}_{12} =  8\mu f_{\mu\tau}^2 
(- \tilde{f}_{e\tau}\omega_{\tau\tau} - \tilde{f}_{e\mu}\omega_{\mu\tau} 
 + \tilde{f}_{e\mu}\tilde{f}_{e\tau}\omega_{e\tau} 
 + \tilde{f}_{e\mu}^2\omega_{e\mu}) I 
= {\cal M}_{21}, \\
{\cal M}_{13} = 8\mu f_{\mu\tau}^2 
(  \tilde{f}_{e\tau}\omega_{\mu\tau} + \tilde{f}_{e\mu}\omega_{\mu\mu} 
 + \tilde{f}_{e\tau}^2\omega_{e\tau} 
 + \tilde{f}_{e\mu}\tilde{f}_{e\tau}\omega_{e\mu}) I 
= {\cal M}_{31}, \\
{\cal M}_{23} = 8\mu f_{\mu\tau}^2 
(  \omega_{\mu\tau} + \tilde{f}_{e\tau}\omega_{e\tau}
 - \tilde{f}_{e\mu}\omega_{e\mu} 
 - \tilde{f}_{e\mu}\tilde{f}_{e\tau}\omega_{ee}) I
={\cal M}_{32}, 
\end{array}
\end{eqnarray}
with the following redefinitions of parameters:
\begin{eqnarray}
{\tilde f}_{e\mu} \equiv \frac{f_{e\mu}}{f_{\mu\tau}},\ \ 
{\tilde f}_{e\tau} \equiv \frac{f_{e\tau}}{f_{\mu\tau}},\ \ 
\omega_{ab} \equiv m_a y_{ab}^* m_b  .
\end{eqnarray}
It is clear that the mass matrix in Eq. (\ref{eq:Mzb}) always has a 
zero-eigenvalue because of the vanishing determinant of $F_{ab}$.
Although all three active neutrinos should have non-zero masses if 
we take into account the higher order loop contributions, we ignore 
these contributions in this Letter.

Furthermore, we can embed the terms associated with the electron mass into 
$\omega_{\tau\tau}$, $\omega_{\mu\tau}$ and $\omega_{\mu\mu}$ terms, 
such that
\begin{eqnarray}
\omega_{\tau\tau}^{'} &\equiv& 
 \omega_{\tau\tau} - 2\tilde{f}_{e\mu}\omega_{e\tau} 
  + \tilde{f}_{e\mu}^2\omega_{ee},
 \nonumber \\
\omega_{\mu\tau}^{'} &\equiv&
 \omega_{\mu\tau} + \tilde{f}_{e\tau}\omega_{e\tau}
 - \tilde{f}_{e\mu}\omega_{e\mu} 
 - \tilde{f}_{e\mu}\tilde{f}_{e\tau}\omega_{ee}, \\
\omega_{\mu\mu}^{'} &\equiv&
 \omega_{\mu\mu} + 2\tilde{f}_{e\tau}\omega_{e\mu} 
 + \tilde{f}_{e\tau}^2\omega_{ee}\nonumber .
\end{eqnarray}
Then, we obtain the following simplified mass matrix
\begin{eqnarray}
{\cal M}_{ab} = 8\mu f_{\mu\tau}^2 \omega_{\mu\mu}^{'} I M_{ab}
\end{eqnarray}
with
\begin{eqnarray}
M_{ab} = 
\left(\begin{array}{ccccc}
 -\tilde{f}_{e\tau}^2 {\tilde \omega}_{\tau\tau} 
 - 2\tilde{f}_{e\mu}\tilde{f}_{e\tau}{\tilde \omega}_{\mu\tau} - \tilde{f}_{e\mu}^2
  & &
 - \tilde{f}_{e\tau}{\tilde \omega}_{\tau\tau} 
 - \tilde{f}_{e\mu}{\tilde \omega}_{\mu\tau} & &
   \tilde{f}_{e\tau}{\tilde \omega}_{\mu\tau} + \tilde{f}_{e\mu} \\
 \ * & & -{\tilde \omega}_{\tau\tau} & & {\tilde \omega}_{\mu\tau} \\
 \ * & & * & & -1
 \end{array}\right), \label{eq:Mab}
\end{eqnarray}
where ${\tilde \omega}_{\mu\tau}$ and ${\tilde \omega}_{\tau\tau}$ 
are defined as
\begin{eqnarray}
{\tilde \omega}_{\mu\tau} 
  = \frac{\omega_{\mu\tau}^{'}}{\omega_{\mu\mu}^{'}},\ \ 
{\tilde \omega}_{\tau\tau} 
  = \frac{\omega_{\tau\tau}^{'}}{\omega_{\mu\mu}^{'}}\ .
\end{eqnarray}

As partially discussed in Ref. \cite{babu-mace}, we can represent 
$\tilde{f}_{e\mu}, \tilde{f}_{e\tau}, \tilde{\omega}_{\mu\tau}$ and
 $\tilde{\omega}_{\tau\tau}$
in terms of the neutrino mass ratios, mixing angles and CP violating phases 
in the PMNS matrix, parametrized by 
\begin{eqnarray}
U_{PMNS}= 
\left(\begin{array}{ccc}
 1 & 0 & 0 \\
 0 &  c_{23} & s_{23} \\
 0 & -s_{23} & c_{23}
\end{array}\right)
\left(\begin{array}{ccc}
 c_{13} & 0 & s_{13}\ e^{-i\delta} \\
 0 &  1 & 0 \\
 -s_{13}\ e^{i\delta} & 0 & c_{13}
\end{array}\right)
\left(\begin{array}{ccc}
  c_{12} & s_{12} & 0 \\
 -s_{12} &  c_{12} & 0 \\
 0 & 0 & 1
\end{array}\right)
\left(\begin{array}{ccc}
 1 & 0 & 0 \\
 0 &  e^{i \gamma/2} & 0 \\
 0 & 0 & 1
\end{array}\right), \label{eq:Vmns}
\end{eqnarray}
where  $\delta$ and $\gamma$ are the Dirac and Majorana CP phase, respectively, 
and $s_{ij}(c_{ij}) = \sin\theta_{ij}(\cos\theta_{ij}) \ge 0$.
Since we consider the diagonal basis of the charged leptons, the Majorana 
mass matrix is diagonalized by the PMNS matrix, such that 
$U^T_{PMNS}\ {\cal M}\ U_{PMNS}=\diag(m_1, m_2, m_3)$.
In the case of the normal mass hierarchy, the four parameters 
are described as
\begin{eqnarray}
&&\tilde{f}_{e\mu} = 
  \frac{s_{12}}{c_{12}}\frac{s_{23}}{c_{13}} 
- \frac{s_{13}}{c_{13}}c_{23}\ e^{i\delta},
\nonumber \\
&&\tilde{f}_{e\tau} = 
  \frac{s_{12}}{c_{12}}\frac{c_{23}}{c_{13}} 
+ \frac{s_{13}}{c_{13}} s_{23}\ e^{i\delta},
\nonumber \\
&&\tilde{\omega}_{\mu\tau} = 
-\frac{ c_{13}^2 s_{23}c_{23} }
     { c_{13}^2 c_{23}^2 
     + r_{2/3} (s_{12}s_{13}c_{23}\ e^{-i\delta} + c_{12}s_{23} )^2 e^{-i\gamma} } 
\label{eq:exEq-NH} \\
&&\hspace{1.2cm}
-\frac{ r_{2/3} (s_{12}s_{13}c_{23}\ e^{-i\delta} + c_{12}s_{23})
            (s_{12}s_{13}s_{23}\ e^{-i\delta} - c_{12}c_{23}) e^{-i\gamma} }
      { c_{13}^2 c_{23}^2 
      + r_{2/3} (s_{12}s_{13}c_{23}\ e^{-i\delta} + c_{12}s_{23} )^2 e^{-i\gamma} },
\nonumber \\
&&\tilde{\omega}_{\tau\tau} = 
\frac{ c_{13}^2 s_{23}^2 
     + r_{2/3} (s_{12}s_{13}s_{23}\ e^{-i\delta} - c_{12}c_{23})^2 e^{-i\gamma} }
     { c_{13}^2 c_{23}^2 
     + r_{2/3} (s_{12}s_{13}c_{23}\ e^{-i\delta} + c_{12}s_{23} )^2 e^{-i\gamma} },
\nonumber
\end{eqnarray}
with $r_{2/3}=m_2 / m_3$, while for the inverted one
\begin{eqnarray}
&&\tilde{f}_{e\mu} =  c_{23}\frac{c_{13}}{s_{13}}\ e^{i\delta} ,
\nonumber \\
&&\tilde{f}_{e\tau} = -s_{23}\frac{c_{13}}{s_{13}}\ e^{i\delta} ,
\nonumber \\
&&\tilde{\omega}_{\mu\tau} = 
-\frac{ r_{2/1} (s_{12}s_{13}c_{23}\ e^{-i\delta} + c_{12}s_{23})
            (s_{12}s_{13}s_{23}\ e^{-i\delta} - c_{12}c_{23}) e^{-i\gamma} }
      { r_{2/1} (s_{12}s_{13}c_{23}\ e^{-i\delta} + c_{12}s_{23})^2 e^{-i\gamma} 
      + (c_{12}s_{13}c_{23}\ e^{-i\delta} - s_{12}s_{23} )^2 }
\label{eq:exEq-IH} \\
&&\hspace{1.2cm}
-\frac{ (c_{12}s_{13}c_{23}\ e^{-i\delta} - s_{12}s_{23})
        (c_{12}s_{13}s_{23}\ e^{-i\delta} + s_{12}c_{23}) }
      { r_{2/1} (s_{12}s_{13}c_{23}\ e^{-i\delta} + c_{12}s_{23})^2 e^{-i\gamma} 
      + (c_{12}s_{13}c_{23}\ e^{-i\delta} - s_{12}s_{23} )^2 } ,
\nonumber \\
&&\tilde{\omega}_{\tau\tau} = 
\frac{ r_{2/1} (s_{12}s_{13}s_{23}\ e^{-i\delta} - c_{12}c_{23})^2 e^{-i\gamma} 
     + (c_{12}s_{13}s_{23}\ e^{-i\delta} + s_{12}c_{23})^2 }
     { r_{2/1} (s_{12}s_{13}c_{23}\ e^{-i\delta} + c_{12}s_{23})^2 e^{-i\gamma} 
     + (c_{12}s_{13}c_{23}\ e^{-i\delta} - s_{12}s_{23} )^2 } ,
\nonumber
\end{eqnarray}
with $r_{2/1}=m_2 / m_1$.
From the first two equations in Eq. (\ref{eq:exEq-NH}), one can see that 
$\tilde{f}_{e\mu}$ will be close to $\tilde{f}_{e\tau}$ in the 
limit of $\theta_{13} \rightarrow 0$ and $\theta_{23} \rightarrow \pi/4$.
This fact turns out to be one of the origins of the $\mu\hif\tau$ symmetry 
as shown in the next section.
On the other hands, $\tilde{f}_{e\mu}$ and $\tilde{f}_{e\tau}$ 
in Eq. (\ref{eq:exEq-IH}) always have an opposite sign.
This indicates that the inverted case cannot be consistent with 
the $\mu\hif\tau$ symmetry
\footnote{
There is actually a special case in which $f_{\mu\tau}=0$.
However, once we force the neutrino mass matrix to be $\mu\hif\tau$ symmetric, 
the theory suffers from the dangerous lepton flavor violating processes, 
such as $\tau\rightarrow \mu\gamma$.
}.

\section{$\mu \hif \tau$ symmetric limit and deviation}
In this section, we investigate the Zee-Babu model 
by considering
the $\mu\hif\tau$ symmetric type of the  matrix in  Eq. (\ref{eq:Mab}) as follows
\begin{eqnarray}
M^{\mu\tau}=
\left(\begin{array}{ccc}
  A & -B & B \\
 -B &  C & D \\
  B &  D & C
\end{array}\right),
\end{eqnarray}
which can be diagonalized by the PMNS matrix in Eq. (\ref{eq:Vmns}) 
with $\theta_{23}=\pi/4$ and $\theta_{13}=0$, 
where $A,\ B,\ C$ and $D$ are complex values in general.
Note that for the matrix in Eq. (\ref{eq:Mab}) there are only two possible
$\mu\hif\tau$ symmetric limits: 
(i) $ \tilde{\omega}_{\mu\tau} = \tilde{\omega}_{\tau\tau} = 1$
$( \omega_{\mu\mu}^{'} = \omega_{\mu\tau}^{'} = \omega_{\tau\tau}^{'} )$ and 
(ii) $\tilde{\omega}_{\tau\tau}=1$ and $\tilde{f}_{e\tau}=\tilde{f}_{e\mu}$
$( \omega_{\mu\mu}^{'} = \omega_{\tau\tau}^{'}$ and $f_{e\mu}=f_{e\tau} )$.
However, the former condition results in $m_1=m_3=0$ or $m_2=m_3=0$, 
which must be largely broken in order to fit the experimental data.
Thus, we will focus on only the latter one.

\subsection{Normal mass hierarchy}
In the $\mu\hif\tau$ symmetric limit, the matrix $M_{ab}$ 
in Eq. (\ref{eq:Mab}) becomes 
\begin{eqnarray}
M_{ab} = 
\left(\begin{array}{ccccc}
 -2\tilde{f}_{e\mu}^2(1+\tilde{\omega}_{\mu\tau}) & & 
 - \tilde{f}_{e\mu}(1+\tilde{\omega}_{\mu\tau}) & & 
   \tilde{f}_{e\mu}(1+\tilde{\omega}_{\mu\tau}) \\
 \ * & & -1 & & \tilde{\omega}_{\mu\tau} \\
 \ * & & * & & -1
 \end{array}\right)
\label{eq:Mmt-2}
\end{eqnarray}
and three mixing angles are given by
\begin{eqnarray}
\theta_{23}=\frac{\pi}{4},\ \ \theta_{13}=0,\ \ 
\tan 2\theta_{12}=
 \frac{2\sqrt{2}\tilde{f}_{e\mu}}{1 - 2\tilde{f}_{e\mu}^2} .
\end{eqnarray}
The three eigenvalues are found to be  
\begin{eqnarray}
&&\lambda_1= 
\left| 
  ( 2\tilde{f}_{e\mu}^2 c_{12}^2 
  - 2\sqrt{2}\tilde{f}_{e\mu} s_{12}c_{12} + s_{12}^2)
  ( \tilde{\omega}_{\mu\tau} + 1) 
\right|, \nonumber\\
&&\lambda_2= 
\left| 
  ( 2\tilde{f}_{e\mu}^2 s_{12}^2 
  + 2\sqrt{2}\tilde{f}_{e\mu} s_{12}c_{12} + c_{12}^2)
  ( \tilde{\omega}_{\mu\tau} + 1) 
\right|, \\
&&\lambda_3= 
 | \tilde{\omega}_{\mu\tau} - 1 |, \nonumber
\end{eqnarray}
where either $\lambda_1$ or $\lambda_2$ always vanishes.
Hence, this limit is only consistent with the normal mass hierarchy case.
For example, the exact TB mixing is obtained from 
$\tilde{f}_{e\mu}=1/2$ \footnote{
Although $\tilde{f}_{e\mu}=-1$ also implies the exact TB mixing, 
it leads to a vanishing $\lambda_2$ at the same time.
}, 
while the central value of the mass ratio, which is $m_2/m_3\simeq0.176$, 
corresponds to $\tilde{\omega}_{\mu\tau}\simeq -1.27$.
Moreover, in order to fit all central values in Eq. (\ref{eq:best}), 
we need to deviate from the $\mu\hif\tau$ symmetric limit and it can be realized
with the following data set:
\begin{eqnarray}
&&\tilde{f}_{e\mu} \simeq 0.47 - 0.07e^{i(0\ \hif\ 2\pi)},\ \ 
  \tilde{f}_{e\tau} \simeq 0.51 + 0.07e^{i(0\ \hif\ 2\pi)},
  \nonumber \\
&&\tilde{\omega}_{\tau\tau} \simeq
  0.85+(0.00\ \hif\ 0.06)e^{i(0\ \hif\ 2\pi)},
  \label{eq:best-NH} \\
&&\tilde{\omega}_{\mu\tau} \simeq
  -0.95 + (0.19\ \hif\ 0.23)e^{i(0\ \hif\ 2\pi)},
  \nonumber
\end{eqnarray}
where we have varied $\delta$ and $\gamma$ from $0$ to $2\pi$  
based on Eq. (\ref{eq:exEq-NH}).
Although the $\mu\hif\tau$ conditions are no longer exact, they 
remain as good approximations, i.e., $\tilde{\omega}_{\tau\tau} \simeq 1$ and 
$\tilde{f}_{e\tau} \simeq \tilde{f}_{e\mu}$.
Therefore, we conclude that there probably exists the $\mu\hif\tau$ symmetry 
behind the TB pattern of neutrino mixings in the case of the normal mass hierarchy.

\subsection{Inverted mass hierarchy}
As mentioned 
in the previous subsection, the $\mu\hif\tau$ condition has to be largely 
broken in the case of the inverted mass hierarchy.
For instance, the central values in Eq. (\ref{eq:best}) can be obtained from 
\begin{eqnarray}
&&\tilde{f}_{e\mu} \simeq 7.49 e^{i(0\ \hif\ 2\pi)},\ \ 
  \tilde{f}_{e\tau} \simeq - 6.94 e^{i(0\ \hif\ 2\pi)},\nonumber \\ 
&&\tilde{\omega}_{\tau\tau} \simeq
  2.00 + (0.00\ \hif\ 1.64)e^{i(0\ \hif\ 2\pi)},
 \label{eq:best-IH} \\
&&\tilde{\omega}_{\mu\tau} \simeq 
  1.52 + (0.00\ \hif\ 0.78)e^{i(0\ \hif\ 2\pi)},
  \nonumber
\end{eqnarray}
where we have varied $\delta$ and $\gamma$ from $0$ to $2\pi$ 
based on Eq. (\ref{eq:exEq-IH}).
However, contrary to the normal hierarchy case, it is difficult to find out 
possible remnants of the $\mu\hif\tau$ symmetry from Eq. (\ref{eq:best-IH}), 
i.e., $\tilde{f}_{e\mu} \neq \tilde{f}_{e\tau}$ and 
$\tilde{\omega}_{\tau\tau} \neq 1$.
This suggests that, in the inverted hierarchy case, the TB mixing may 
just be an accidental result due to the suitable parameter tunings. 

\section{Froggatt-Nielsen-like $Z_5$ model}
\begin{table}[t]
\begin{center}
\begin{tabular}{|c||c|c|c|c|c|c|}\hline
       & $L$ & $\ell_R$ & $H$ & $h^+$ & $k^{++}$ & $\phi$ \\ \hline
 $SU(2)_L$ & $2$ &  $1$ & $2$ & $1$ & $1$ & $1$ \\ \hline
 $U(1)_Y$ & $-1$ &  $-2$ & $1$ & $2$ & $4$ & $0$ \\ \hline
 $Z_5$ & $(2,0,1)$ &  $(2,0,1)$ & $0$ & $0$ & $0$ & $-1$ \\ \hline
\end{tabular}
\end{center}
\caption{A particle content and charge assignments}
\label{tab:z5}
\end{table}
In the previous section, we have obtained the conditions:
$\tilde{f}_{e\tau}=\tilde{f}_{e\mu}$ and $\tilde{\omega}_{\tau\tau}=1$ 
to derive the $\mu\hif\tau$ symmetric matrix.
The former condition is easy to achieve by imposing a permutation 
symmetry, whereas the latter one may not be because 
$\tilde{\omega}_{\tau\tau}$ includes not only Yukawa couplings 
but also the charged lepton masses.
For instance, if we ignore the electron mass, the condition becomes 
$m_\tau^2 y_{\tau\tau}^* = m_\mu^2 y_{\mu\mu}^* $, and it requires 
a hierarchy between Yukawa couplings rather than a permutation relation.
To naturally realize the $\mu\hif\tau$ conditions for the normal mass hierarchy 
case, we adopt the scheme of the Froggatt-Nielsen mechanism \cite{FN} and show 
a specific model based on an $Z_5$ symmetry. 
The charge assignments of the particles under the symmetries are summarized 
in Table \ref{tab:z5}.
In this model, we introduce a gauge singlet scalar $\phi$ with the charge $-1$ 
under $Z_5$ and consider the higher dimensional operators.
Because of the $Z_5$ symmetry, at the leading order, the Yukawa matrices 
$F_{ab}$ and $Y_{ab}$ in Eq. (\ref{eq:zbL}) turn out to be
\begin{eqnarray}
F_{ab}=
 \left(\begin{array}{ccc}
 0 & f_{e\mu}\lambda^2 & f_{e\tau}\lambda^2 \\
 -f_{e\mu}\lambda^2 & 0 & f_{\mu\tau}\lambda \\
 -f_{e\tau}\lambda^2 & -f_{\mu\tau}\lambda & 0
 \end{array}\right),
\hspace{1cm}
Y_{ab}=
 \left(\begin{array}{ccc}
 y_{ee}\lambda & y_{e\mu}\lambda^2 & y_{e\tau}\lambda^2 \\
 \ * & y_{\mu\mu} & y_{\mu\tau}\lambda \\
 \ * & \ * & y_{\tau\tau}\lambda^2
 \end{array}\right),
\end{eqnarray}
where $\lambda=<\phi>/\Lambda$ is the suppression factor of the higher 
dimensional operators with the typical energy scale of the $Z_5$ symmetry, 
$\Lambda$.
Note that due to the symmetry, the charged lepton mass matrix is diagonal 
up to the leading order.
We remark that to simplify our discussion, 
we have assumed that the terms like $LHLH\phi^{(*)n}$  are strongly
suppressed by an extremely-high energy scale.
It is easy to see that if we assume
\begin{eqnarray}
\lambda = \frac{m_\mu}{m_\tau} \simeq 0.06,\ \ 
f_{e\tau}=f_{e\mu},\ \ 
y_{\mu\mu}=y_{\tau\tau} \label{eq:z5-mt}
\end{eqnarray}
and $m_e = 0$, we obtain the $\mu\hif\tau$ symmetric Majorana neutrino 
mass matrix, given by 
\begin{eqnarray}
{\cal M}_{ab} = 8\mu f_{\mu\tau}^2 \lambda^2 \omega_{\mu\mu}
 \left(\begin{array}{ccccc}
 -2\tilde{f}_{e\mu}^2 \lambda^2 (1+\tilde{\omega}_{\mu\tau}) & & 
 -\tilde{f}_{e\mu} \lambda (1+\tilde{\omega}_{\mu\tau}) & & 
  \tilde{f}_{e\mu} \lambda (1+\tilde{\omega}_{\mu\tau}) \\
 \ * & & -1 & & \tilde{\omega}_{\mu\tau} \\
 \ * & & * & & -1
\end{array}\right) I ,
\end{eqnarray}
where $\tilde{\omega}_{\mu\tau}=y_{\mu\tau}^* / y_{\mu\mu}^* $ and 
$\omega_{\mu\mu}=m_\mu^2 y_{\mu\mu}^*$.

\begin{table}
\begin{center}
\begin{tabular}{ccc}\hline
 $\mu\rightarrow e\gamma$ & & Br. $< 1.2\times 10^{-11}$ \\ \hline 
 $\mu\rightarrow e^+ e^- e^- $ &\ & Br. $< 1.0\times 10^{-12}$ \\ \hline 
 $\tau\rightarrow \mu^+ \mu^- \mu^-$ &\ & Br. $< 3.2\times 10^{-8}$ \\ \hline
\end{tabular}
\end{center}
\caption{Lepton flavor violating processes and their experimental bounds 
used in our calculation.}
\label{tab:lfv}
\end{table}
In the rest of this section, we discuss several experimental constraints 
including some non-standard lepton flavor violating processes.
(See Table \ref{tab:lfv}.)
As discussed in Sec. III, the realistic neutrino mixings requires
\begin{eqnarray}
 f_{e\mu} \simeq f_{e\tau} \simeq \frac{f_{\mu\tau}}{2\lambda} .
\end{eqnarray}
In this case, the strongest constraint on the $f_{ab}$ couplings comes 
from the $\mu\rightarrow e\gamma$ process.
Combined with the neutrino mixing data, we get the lower bounds on 
$f_{\mu\tau}$ and $y_{\mu\mu}$:
\begin{eqnarray}
 &&f_{\mu\tau}\lambda > 0.008, \\
 &&y_{\mu\mu} > 0.13 ,
\end{eqnarray}
and the allowed range for the singly charged scalar mass:
\begin{eqnarray}
10^2\ \gev < M_h < 10^4\ \gev
\end{eqnarray}
as discussed in Ref. \cite{babu-mace}.
Moreover, since it may be natural to take all $y_{ab}$ to be the 
same order, we can estimate the branching ratios of 
$\ell_a^- \rightarrow \ell_b^+ \ell_c^- \ell_c^-$ 
mediated by the doubly charged scalar.
The stringent constraint comes form either $\tau\rightarrow 3\mu$ or
$\mu \rightarrow 3e$, given by
\begin{eqnarray}
 \frac{|y_{ab}|^2}{M_k^2} < 10^{-7}\ \gev^{-2}.
\end{eqnarray}
It is clear that these processes can be accessible in the near future 
with the $\tev$ scale doubly charged scalar if $y_{ab} \simeq {\cal O}(0.1)$.

\section{Summary}
We have investigated the Zee-Babu model and tried to find out a possible 
flavor symmetry behind the TB neutrino mixing matrix.
We have found that there probably exists the $\mu\hif\tau$ symmetry in the 
normal neutrino mass hierarchy case, but the TB mixing may be accidental 
in the case of the inverted one.
We have also attempted to derive the $\mu\hif\tau$ symmetric neutrino mass matrix 
with a Froggatt-Nielsen-like $Z_5$ symmetry and estimated several constraints 
coming from lepton flavor violating processes.
\\

\noindent {\bf Acknowledgments}

This work is supported in part by the National Science Council of ROC under 
Grants Nos. NSC-95-2112-M-007-059-MY3 and NSC-98-2112-M-007-008-MY3 
and by the Boost Program of NTHU.


\begin{thebibliography}{3}
\bibitem{PMNS}
Z. Maki, M. Nakagawa and S. Sakata, {\em Prog. Theor. Phys.} {\bf 28}, 870 (1962);
B. Pontecorvo, {\em Zh. Eksp. Teor. Fiz.} {\bf 53}, 1717 (1967); 
{\em Sov. Phys. JETP} {\bf 26}, 984(1968).

\bibitem{gfit}
M. C. Gonzalez-Garcia, M. Maltoni and J. Salvado, 
{\em JHEP} {\bf 1004}, 056 (2010). 

\bibitem{TB}
P. F. Harrison, D. H. Perkins and W. G. Scott, 
{\em Phys. Lett. B} {\bf 530}, 167 (2002);
P. F. Harrison and W. G. Scott, {\em Phys. Lett. B} {\bf 535}, 163 (2002).

\bibitem{type1}
P. Minkowski, {\em Phys. Lett. B} {\bf 67}, 421 (1977);
T. Yanagida, in {\em Proceedings of the Workshop on Unified Theories
and Baryon Number in the Universe}, eds.\ O.~Sawada and A.~Sugamoto
(KEK report 79-18, 1979); 
M. Gell-Mann, P. Ramond and R. Slansky, 
in {\em Supergravity}, eds.\ P. Van Nieuwenhuizen and D. Z. Freedman
(North Holland, Amsterdam, 1979);
R. N. Mohapatra and G. Senjanovic, {\em Phys. Rev. Lett.} {\bf 44}, 912 (1980).

\bibitem{type2}
W. Konetschny and W. Kummer, {\em Phys. Lett. B} {\bf 70}, 433 (1977); 
J. Schechter and J. W. F. Valle, {\em Phys. Rev. D} {\bf 22}, 2227 (1980);
T. P. Cheng and L. F. Li, {\em Phys. Rev. D} {\bf 22}, 2860 (1980);
G. Lazarides, Q. Shafi and C. Wetterich, {\em Nucl. Phys. B} {\bf 181}, 287 (1981).

\bibitem{type3}
R. Foot, H. Lew, X. G. He and G. C. Joshi, {\em Z. Phys. C} {\bf 44}, 441 (1989);
E. Ma, {\em Phys. Rev. Lett} {\bf 81}, 1171 (1998);
E. Ma and D. P. Roy, {\em Nucl. Phys. B} {\bf 644}, 290 (2002).

\bibitem{dim5}
S. Weinberg, {\em Phys. Rev. Lett.} {\bf 43}, 1566 (1979).

\bibitem{mu-tau}
T. Fukuyama and H. Nishiura, 
in {\it Proceedings of 1997 Shizuoka Workshop on Masses and Mixings of Quarks and Leptons}, hep-ph/9702253;
G. Altarelli and F. Feruglio, {\em Phys. Lett. B} {\bf 439}, 112 (1998).

\bibitem{A4}
E. Ma, {\em Phys. Rev. D} {\bf 73}, 057304 (2006);
G. Altarelli and F. Feruglio, {\em Nucl. Phys. B} {\bf 741}, 215 (2006).

\bibitem{S4}
C. S. Lam, {\em Phys. Lett. B} {\bf 656}, 193 (2007); 
{\em Phys. Rev. Lett.} {\bf 101}, 121602 (2008); 
{\em Phys. Rev. D} {\bf 78}, 073015 (2008). 

\bibitem{tani-rev}
H. Ishimori, T. Kobayashi, H. Ohki, H. Okada, Y. Shimizu and M. Tanimoto,
arXiv:1003.3552 [hep-th] and references therein.

\bibitem{FL}
R. Friedberg and T. D. Lee, {\em High Energy Phys. Nucl. Phys.} {\bf 30}, 591 (2006);
{\em Annal Phys.} {\bf 323}, 1087 (2008);
Z. Z. Xing, H, Zhang, and S. Zhou, {\em Phys. Lett. B} {\bf 641}, 189 (2006);
S. Luo and Z. Z. Xing, {\em Phys. Lett. B} {\bf 646}, 242 (2007);
T. Araki and R. Takahashi, {\em Eur. Phys. J. C} {\bf 63}, 521 (2009);
T. Araki and C. Q. Geng, {\em Phys. Lett. B} {\bf 680}, 343 (2009).

\bibitem{zee}
A. Zee, {\em Phys. Lett. B} {\bf 93}, 389 (1980).

\bibitem{zee-babu}
A. Zee, {\em Nucl. Phys. B} {\bf 264}, 99 (1986);
K. S. Babu, {\em Phys. Lett. B} {\bf 203}, 132 (1988).

\bibitem{variety}
L. M. Krauss, S. Nasri and M. Trodden, {\em Phys. Rev. D} {\bf 67}, 085002 (2003);
E. Ma, {\em Phys. Rev. D} {\bf 73}, 077301 (2006);
C. S. Chen, C. Q. Geng and J. N. Ng, {\em Phys. Rev. D} {\bf 75}, 053004 (2007);
C. S. Chen, C. Q. Geng, J. N. Ng and J. M. S. Wu, {\em JHEP} {\bf 0708}, 022 (2007);
M. Aoki, S. Kanemura and O. Seto, {\em Phys. Rev. Lett.} {\bf 102}, 051805 (2009);
A. Adulpravitchai, M. Lindner, A. Merle and R. N. Mohapatra, 
{\em Phys. Lett. B} {\bf 680}, 476 (2009); 
C. S. Chen and C. Q. Geng, arXiv:1005.2817 [hep-ph].

\bibitem{antiZee}
C. Jarlskog, M. Matsuda, S. Skadhauge and M. Tanimoto, 
{\em Phys. Lett. B} {\bf 449}, 240 (1999);
X. G. He, {\em Eur. Phys. J. C} {\bf 34}, 371 (2004). 

\bibitem{babu-mace}
K. S. Babu and C. Macesanu, {\em Phys. Rev. D} {\bf 67}, 073010 (2003). 

\bibitem{pheno}
D. A. Sierra and M. Hirsch, {\em JHEP} {\bf 0612}, 052 (2006); 
T. Ohlsson, T. Schwetz and H. Zhang, {\em Phys. Lett. B} {\bf 681}, 269 (2009);
M. Aoki, S. Kanemura, T. Shindou and K. Yagyu, arXiv:1005.5159 [hep-ph].


\bibitem{FN}
 C. D. Froggatt and H. B. Nielsen, {\em Nucl. Phys. B} {\bf 147}, 277 (1979).
 
\end{thebibliography}
\end{document}